\documentclass[superscriptaddress, twocolumn,10pt, prb, aps, showpacs]{revtex4-2}

\usepackage{fancyhdr} %
\usepackage[caption=false, position=top,textfont=normalfont,singlelinecheck=off,justification=raggedright]{subfig}
\usepackage{tabularx} %
\usepackage{graphicx} %
\usepackage{float}
\usepackage[table]{xcolor} %
\usepackage{dsfont} %
\usepackage{physics} %

\usepackage{lipsum} %
\usepackage{url}
\usepackage[colorlinks, pdfpagelabels=false, hypertexnames = false]{hyperref} %
\hypersetup{
 breaklinks=true,
 hypertexnames=false,
 colorlinks=true,
 linkcolor={blue},
 citecolor={red},
 urlcolor={blue},
 anchorcolor={black}
}
\usepackage{orcidlink}

\makeatletter
\def\maketitle{
\@author@finish
\title@column\titleblock@produce
\suppressfloats[t]}
\makeatother

\begin{document}

\author{David S. Schlegel\,\orcidlink{0000-0003-2013-1676}}
\email{david.schlegel@epfl.ch}
\affiliation{Laboratory of Theoretical Physics of Nanosystems (LTPN), Institute of Physics, École Polytechnique Fédérale de Lausanne (EPFL), CH-1015 Lausanne, Switzerland}
 \affiliation{Center for Quantum Science and Engineering, École Polytechnique Fédérale de Lausanne (EPFL), CH-1015 Lausanne, Switzerland}
\author{Stefan Kehrein\,\orcidlink{0009-0005-1119-4124}}
\email{stefan.kehrein@theorie.physik.uni-goettingen.de}
\affiliation{Institute for Theoretical Physics, Georg-August-Universit\"at G\"ottingen, Friedrich-Hund-Platz 1, 37077 G\"ottingen, Germany}

\title{Measurement-induced Clock in a Lattice Ring of Non-interacting Electrons}

\begin{abstract}
We examine the emergence of periodicity in a non-interacting steady-state quantum system without external drive inspired by quantum time crystals' spontaneous time-translation symmetry breaking. Specifically, we consider a lattice ring of non-interacting electrons undergoing weak local position measurements. Our analysis uncovers time-periodic structures in steady-state two-time correlation functions, with periodicity linked to the system's group velocity. This study demonstrates a measurement-induced clock mechanism, highlighting periodic behaviors in two-time correlators of a non-equilibrium steady state, contributing to understanding time-periodic phenomena in minimally interactive quantum systems.
\end{abstract}

\date{\today}

\maketitle
\section{Introduction}
Open quantum systems, which explicitly consider the interaction between the system and its environment, have become a major field of study~\cite{Weiss2011, Rivas2012, Breuer2015}, shedding light on fundamental processes such as dephasing, decoherence, and particle loss~\cite{Breuer2015, Macchiavello2001, Schlosshauer2007, Schaller2014, Palma1996}. They play a pivotal role in understanding a wide range of experimental setups in quantum science and technology, such as photonic systems~\cite{Louisell1973, Cohen-Tannoudji1998, Carmichael2008, Szameit2010}, ultracold atoms~\cite{Bloch2008, Dimer2007, Baumann2010, Baumann2011, Brennecke2013}, condensed matter systems~\cite{Leggett1987, Blum1996, Dattagupta2004, Weiss2011, Fitzpatrick2017, Sato2019, McIver2019}, and chemical physics~\cite{May2011, Nitzan2006}. Additionally, they are particularly crucial in quantum computation, where quantum bits are sensitive to decoherence due to environmental interaction~\cite{Nielsen2010}. Open quantum systems also display emergent phenomena, in particular, dissipative phase transitions~\cite{Jin2016, Kessler2012, Diehl2010, Khemani2016, Fitzpatrick2017, Minganti2018} and topological phases~\cite{Dong2016, St-Jean2017, Lu2014, Snizhko2019, Bardyn2013}.

Spontaneous symmetry breaking is a ubiquitous phenomenon observed in nature, underpinning a diverse array of phenomena. This phenomenon is characterized by a situation in which the equations that govern a system exhibit a certain degree of symmetry, but the system selects a solution in a spontaneous manner that breaks this symmetry. This effect can be attributed to the susceptibility of exact symmetric eigenstates to infinitesimally weak perturbations. In particular, the spontaneous breaking of time-translation symmetry, in which a quantum many-body system self-organizes in time and starts spontaneously to undergo a periodic motion, is a key characteristic of quantum \emph{time crystals}~\cite{Wilczek2012}, opening up new avenues for exploring non-equilibrium phases of matter in interacting quantum systems.

Although it has been proven that time-translation symmetry cannot spontaneously be broken in the ground state of quantum systems with short-range interactions~\cite{Bruno2013a, Watanabe2015, Kozin2019}, the formation of \emph{discrete} time-crystals in driven out-of-equilibrium systems can be studied theoretically in the framework of \emph{Floquet theory}~\cite{Sacha2015, Else2016a, Khemani2016, Syrwid2017, Moessner2017a, Khemani2017, Ho2017}.
Recently, discrete time crystals have also been observed experimentally~\cite{Zhang2017, Choi2017, Rovny2018, Pal2018, Sullivan2018, Kyprianidis2021, Taheri2022}. In these systems, many-body interactions play a fundamental role in the formation of periodic motion with a period different from the external drive~\cite{Sacha2018, Choi2017, Else2016a, Ho2017, Moessner2017a, Yao2017}.
Furthermore, systems with long-range interactions have been theoretically analyzed that exhibit spontaneous time-translation symmetry breaking in the ground state~\cite{Kozin2019}.
Tremendous efforts have been made to understand time crystallinity in open quantum systems~\cite{Lazarides2017, Iemini2018, Lazarides2019, Seibold2019, Riera-Campeny2019}, where no-go theorems for time crystals in the ground state of closed systems do not apply~\cite{Syrwid2017}.

In recent studies, repeatedly measuring a quantum system revealed various measurement-induced phenomena, such as measurement-induced quantum jumps~\cite{BauerJournalofPhysicsA2015}, measurement-induced nonlocality~\cite{LuoPRL2011, XiPRA2012, HuAnnalsofPhysics2012, HuNJP2015}, and measurement-induced phase transitions~\cite{SkinnerPRX2019, TangPRR2020, ChoiPRL2020, JianPRL2021, BuchholdPRX2021, MinatoPRL2022, LiPRL2023} in particular.
In continuously monitored many-body systems, measurement-induced phase transitions in the entanglement entropy have been studied~\cite{TurkeshiPRB2021, TurkeshiPRB2022, GalarXiv2023}, revealing a rich underlying physics in spin systems.

In many theoretical studies of time crystals and systems of measurement-induced phenomena, many-body interactions play a fundamental role.
An interesting and less explored question is: Can time-periodicity be observed in non-interacting but solely dissipative systems in their steady state, without the need for external drives or many-body interactions, in which only the interaction with the environment in the form of continuous measurements induces periodic motion in the system?
The study of transport in monitored non-interacting Bose-Hubbard chains~\cite{TurkeshiarXiv2023} displays a profound difference between the average and \emph{typical} properties of the system. This suggests the existence of non-trivial spatial density and current profiles that can reveal periodicity in two-time correlations in the steady state of a system with closed boundary conditions.

In this work, we answer the question above in the affirmative by realizing a \emph{measurement-induced clock} in an open quantum system out of equilibrium. We propose a toy model consisting of a ring of non-interacting electrons on a lattice, described by a tight-binding model. The system is subject to local dissipators, resembling continuous quadi-local position measurements.
Importantly, contrary to Ref~\cite{TurkeshiarXiv2023}, the local position measurements modelled by dissipators in our studied system are quasi-local, with a tunable locality parameter, determining the localization of the wave-function.

We analyze the system's dynamics using two approaches: the Lindblad master equation and the quantum trajectory method. We exactly solve the system for uniform dissipation at every lattice site.
We observe long-lived decaying oscillations in two-time correlators with a bimodal probability current distribution in the steady state. Throughout the dynamics, the repeated measurements localize the wave function of individual quantum trajectories in space so that the system remains highly localized throughout its dynamics, reminiscent of soliton dynamics~\cite{Grelu2012}. As a change of direction of the system's wave function occurs at a characteristic time scale, the two-time correlations in the steady-state ultimately fade.
The toy model of a measurement-induced clock explored in this work shows rich dynamical behavior in the absence of many-body interactions and solely induced by measurements from an environment.

We describe the system setup in Sec.~\ref{sec:systemsetup} and detail our results in Sec.~\ref{sec:results}.
Sec.~\ref{sec:conclusions} contains the conclusions and outlook.

\section{System setup}
\label{sec:systemsetup}
\begin{figure*}[t]
\includegraphics[width=\textwidth]{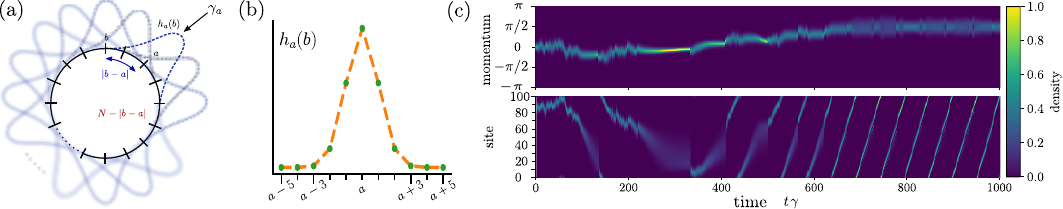}
\caption{(a) Schematic representation of the system setup. At each lattice point $a$, a Lindblad operator acts locally on the system with amplitude function $h_a(b)$, and dissipation rate $\gamma_a$. (b) Amplitude function \mbox{$h_a(b)$} of the Lindblad operator $\hat{D}_a$ for constant lattice point $a$ and $\sigma=1$. (c) Probability density of a sample trajectory as a function of time in quasi-momentum (top) and real space representation (bottom). System parameters: $\gamma_a=\mathrm{t}_\mathrm{hop}$ $\forall a$,  $N=100$, $\sigma=0.1N$; initial state: $\hat{\rho}_0 = \ketbra{k_0}{k_0} = \ketbra{0}{0}$. Due to the measurement-induced interaction, the particle starts moving along the chain in a sustained periodic motion.}
\label{fig:Fig1}
\end{figure*}
Hereinafter, we will analyze a tight-binding model for non-interacting electrons on a lattice with periodic boundary conditions that couple to an environment acting locally on the system, as sketched in Fig.~\ref{fig:Fig1}(a).
With this model, we investigate temporal structures with a particular focus on the steady state.
In order to closely resemble the situation of an experimental setup, we model the system's environment by dissipators that depend on the position of the lattice.
This can be viewed as continuously probing the system by a measurement apparatus that acts locally on the system and thereby induces dissipation.
Such a position-dependent coupling to an environment can generally be described using an effective Lindblad master equation with quasi-local Lindblad operators, describing the reduced system dynamics, as long as the environmental interaction allows for a Markovian approximation.

As the environment is exciting the system, we would intuitively expect the particle to gain momentum, starting from the ground state.
The quasi-local Lindblad operators of the dissipation mechanism would naturally lead to a localization of the wave function in the stochastic dynamics considered.
Hence, this simple model provides an ideal playground for analyzing temporal structures that are induced by the environment acting on the system.

We consider an open quantum system coupled to a Markovian bath \cite{Breuer2015, Schaller2014a}, whose dynamics is governed by the Lindblad master equation
\begin{equation}
    \label{eq:Lindblad_general}
    \dv{\hat{\rho}}{t} = -i \left[\hat{H}, \hat{\rho}\right] + \sum\limits_j \gamma_j \mathcal{D}\left[\hat{L}_j\right](\hat{\rho}).
\end{equation}
Here, $\hat{H}$ is the Hamiltonian and  $\mathcal{D}[\hat{L}_j](\hat{\rho})$ are the dissipators with Lindblad operators $\hat{L}_j$ and dissipation rates $\gamma_j$, with the dissipator defined by \mbox{$\mathcal{D}[\hat{A}](\hat{\rho}) =\hat{A}\hat{\rho}\hat{A}^\dagger - \frac{1}{2} \hat{A}^\dagger\hat{A}\hat{\rho} - \frac{1}{2}\hat{\rho}\hat{A}^\dagger \hat{A}$}.

\subsection{Tight-Binding Hamiltonian}
We consider a system of free spinless non-interacting electrons on a discrete lattice with N sites, periodic boundary conditions, and nearest-neighbor hopping, modeled by a tight-biding Hamiltonian~\cite{Altland2006}:
\begin{equation}\label{eq:Hamiltonian}
\hat{H}_\text{TB} = -\mathrm{t}_\mathrm{hop}\sum\limits_{a=1}^N(\hat{c}_a^\dagger \hat{c}_{a+1} + \hat{c}_{a+1}^\dagger \hat{c}_a),\\
\end{equation} with $\hat{c}_{N+1} = \hat{c}_1$, where $\mathrm{t}_\mathrm{hop}$ is the hopping amplitude. Here, $\hat{c}_a^\dagger$ and $\hat{c}_a$ are the electronic creation and annihilation operators, respectively. The system can be diagonalized in quasi-momentum space, where the Hamiltonian reads:
\begin{equation}
\hat{H}_\text{TB} = \sum\limits_k\varepsilon_k\hat{c}_k^\dagger \hat{c}_k,
\end{equation} with the momentum space annihilation and creation operators $\hat{c}_k$ and $\hat{c}_k^\dagger$.
In this representation, one obtains the dispersion relation~\cite{Ashcroft1976}
\begin{equation}
\varepsilon_k = -2\mathrm{t}_\mathrm{hop}\cos (k\mathrm{a}),
\end{equation} where $\mathrm{a}$ is the lattice constant.
As we deal with periodic boundary conditions, the allowed $k$-values are restricted to the first Brillouin zone $(-\pi/\mathrm{a} \leq k < \pi/\mathrm{a})$, such that $k=\frac{2\pi}{Na}m$, with $m \in \left\{-N/2,-N/2+1,\dots,N/2-1\right\}$. In the following, we will set the lattice spacing $\mathrm{a}=1$ for simplicity. The set of corresponding quasi-momentum states $\{\ket{k}\}$ represent a complete and orthonormal basis of the systems' Hilbert space.

\subsection{Measurement Interaction}
We consider a weak local interaction with the system, resembling continuous measurements of the particle's position around each lattice point.
We model the dissipation induced by this environmental interaction using dissipators with quasi-local Lindblad operators acting on the \emph{position} of the particle around a region of a certain lattice point. We consider Hermitian Lindblad operators $\hat{D}_a$ acting around lattice point $a$:
\begin{equation}\label{eq:D_a_Fourier}
\begin{split}
\hat{D}_a &= \sum\limits_{b=1}^N h_a( b)\hat{c}_b^\dagger \hat{c}_b \\
&=\frac{1}{N} \sum\limits_{k,k'}\hat{c}_{k}^\dagger \hat{c}_{k'}\sum\limits_b h_a( b) e^{\mathrm{i}(k-k')b}.
\end{split}
\end{equation}
Here, $h_a(b)$ is an amplitude function that depends on the distance between the lattice points $a$ and $b$.
The amplitude function $h_a(b)$ of a Lindblad operator $\hat{D}_a$ models the strength and locality of the interaction with the system around a lattice point $a$, depending only on the spatial distance between two lattice points $a$ and $b$.

As a result, measuring the particle will result in the localization of its position. Repeated week measurements of an initial delocalized particle will further localize the particle's wave function.

Due to the periodic boundary conditions, the amplitude function $h_a(b)$ needs to incorporate both distances, $|a-b|$ and $N-|a-b|$.
We consider $h_a(b)$ having a Gaussian form:
\begin{equation}
h_a(b) =\frac{1}{\sqrt{N_h}}\left(e^{-\frac{1}{2}\left(\frac{b-a}{\sigma}\right)^2}\ + e^{-\frac{1}{2}\left(\frac{N-|b-a|}{\sigma}\right)^2}\right),
\end{equation}with the variance $\sigma^2$ characterizing the width of the Gaussian and a normalization constant $N_h$ which is chosen such that:
\begin{equation}
\bra{\psi}\sum\limits_{a=1}^N \hat{D}_a^\dagger \hat{D}_a \ket{\psi} = 1,
\end{equation}for any arbitrary state $\ket{\psi}$. This requirement fixes the normalization:
\begin{equation}\label{eq:normalization_h_a}
\sum\limits_{a=1}^N h_a^2(b) = 1\quad \forall b \in [1,\dots, N]
\end{equation}
The variance $\sigma^2$ describes the broadening of the amplitude function $h_a(b)$ around lattice point $a$ and is thus directly related to the uncertainty of the Lindblad operator $\hat{D}_a$: For larger values of $\sigma$, the operators $\hat{D}_a$ will induce a larger delocalization on the lattice and correspondingly a stronger localization in quasi-momentum space. The amplitude function $h_a(b)$ is schematically depicted in Fig.~\ref{fig:Fig1}(b).
\par
To express $\hat{D}_a$ in the quasi-momentum basis, we compute the discrete Fourier transform of $h_a(b)$:
\begin{equation} 
\mathcal{F}[h_a(b)](q) = \frac{1}{\sqrt{N}}\sum\limits_{b=1}^N h_a(b) e^{\mathrm{i}qb} \equiv \eta(q)e^{-iqa},
\end{equation}
where we have defined $\eta(q) \equiv \mathcal{F}[h_0(b)](q)$.
Eq. (\ref{eq:D_a_Fourier}) then becomes
\begin{equation}
\hat{D}_a =\frac{1}{\sqrt{N}} \sum\limits_{k,k'}\hat{c}_{k}^\dagger \hat{c}_{k'} \eta(k-k') e^{-\mathrm{i}a(k-k')}.
\end{equation} 

With the system Hamiltonian $\hat{H}_\text{TB}$ and the Lindblad operators $\hat{D}_a$, the Lindblad master equation reads
\begin{equation}\label{eq:Linbdblad_a}
\begin{split}
\frac{\text{d}}{\text{d}t} \hat{\rho}
&=-\mathrm{i}[\hat{H}_\text{TB},\hat{\rho}] + \sum\limits_{a=1}^N\gamma_a\mathcal{D}[\hat{D}_a](\hat{\rho}).
\end{split}
\end{equation}
The rates $\gamma_a$ are real positive coefficients representing the \emph{dampening} amplitude of the corresponding damping channel for the Lindblad operators $\hat{D}_a$ each acting locally a region around a lattice point $a$.

\section{Results}
\label{sec:results}
We now turn to results obtained from direct numerical integration of the Lindblad master equation and obtained from the quantum trajectory method.
In Fig.~\ref{fig:Fig1}(c), we show a sample quantum trajectory for an initially fully delocalized state. We observe the emergence of a sustained periodic motion and localization of wave function induced by dissipation.

\subsection{Equations of Motion}
We can express the Lindblad master equation as a super-operator $\mathcal{L}$ acting on the system's density matrix $\hat{\rho}$:
\begin{equation}
\frac{\text{d}}{\text{d}t} \rho_{kk'} \ketbra{k}{k'}= \sum\limits_{\alpha\beta} \mathcal{L}_{ kk'\alpha\beta}\rho_{\alpha\beta}\ketbra{\alpha}{\beta}.
\end{equation}
Explicitly computing the matrix elements of the Liouvillian leads to the following explicit form of the Lindblad master equation [See Appendix \ref{app:LiouvillianEOM} for detailed derivation]:
\begin{equation}\label{eq:Lindblad_final}
\begin{split}
&\mathcal{L}[\rho_{k k'}] = -\mathrm{i}\left( \varepsilon_k - \varepsilon_{k'} \right) \rho_{kk'} \\&+ \frac{1}{N}\sum\limits_a \gamma_a  \left\{ \sum\limits_{\alpha,\beta}\left[\eta(\alpha-k)\eta(k'-\beta) e^{-\mathrm{i}a(\alpha-\beta+k'-k)}\rho_{\alpha\beta}\right.\right.\\
&\qquad-\frac{1}{2} \eta(\alpha-\beta)\eta(\beta-k)e^{-\mathrm{i}a(\alpha-k)}\rho_{\alpha k'} \\
&\left. \left.\qquad- \frac{1}{2}\eta(k'-\alpha)\eta(\alpha-\beta)e^{-\mathrm{i}a(k'-\beta)}\rho_{k \beta}\right]\right\}.
\end{split}
\end{equation}

In the case of all Lindblad operators acting on the system with equal dissipation rates, i.e. $\gamma_a = \gamma = \text{const.}$, the above equation reduces to:

\begin{equation}
\label{eq:Lindbladequaldissipation}
\begin{split}
\mathcal{L}[\rho_{k k'}] &= \left[-\mathrm{i}\left( \varepsilon_k - \varepsilon_{k'} \right)-\gamma\right]\rho_{k k'} \\
&+ \gamma\sum\limits_{\alpha\beta}\eta(\alpha-k)\eta(k'-\beta)\delta_{\alpha-\beta, k-k'}\rho_{\alpha\beta}.
\end{split}
\end{equation}

In the following, we consider the case in which all Lindblad operators $\hat{D}_a$ act on the system with equal dissipation rates $\gamma_a=\gamma$. This resembles the physical situation in which each lattice point interacts equally with its surrounding environment.

\begin{figure}[tb]
	\centering
	\includegraphics[width=\columnwidth]{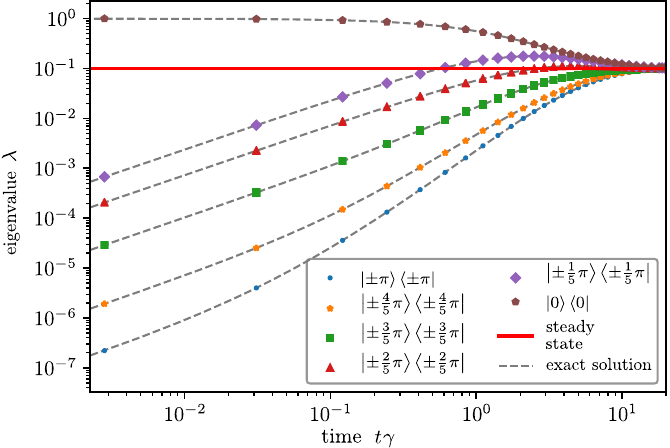}
	\caption{Diagonal matrix elements of the density matrix $\hat{\rho}$ as a function of time for $N=10$ sites. The markers indicate simulation results from direct numerical integration. Here, the system is initialized with $\hat{\rho} = \ketbra{k_0}{k_0} = \ketbra{0}{0}$.}
	\label{fig:0303diagonalelementslogwithanalytic}
\end{figure}

With an initially mixed density $\hat{\rho}_0 = \sum_k \rho_{k k}\ketbra{k}{k}$, the time-evolved density matrix remains diagonal and real at all times. The time-evolution of the diagonal elements is then given by [See Appendix~\ref{app:exactresult} for detailed derivation]
\begin{equation}\label{eq:exact_result}
\frac{\text{d}}{\text{d}t} \rho_{kk} = \gamma\left(\sum\limits_\alpha \eta^2(\alpha-k)\rho_{\alpha\alpha}-\rho_{kk}\right).
\end{equation}
The above equation is a coupled linear differential equation and can be solved by diagonalizing $\mathcal{L}_{\alpha k} = \eta^2(\alpha-k)-\delta_{\alpha k}$.
Together with the numerical result, the exact solution to the above equation is shown in Fig. \ref{fig:0303diagonalelementslogwithanalytic} for an initial pure state with zero-quasi-momentum ${\hat{\rho}} = \ketbra{k=0}{k=0}$.
The resulting steady state is an equilibrium state corresponding to infinite temperature where each quasi-momentum state is equally occupied.

\subsection{Two-Time Correlations}
\begin{figure}[b]
	\centering
	\includegraphics[width=\columnwidth]{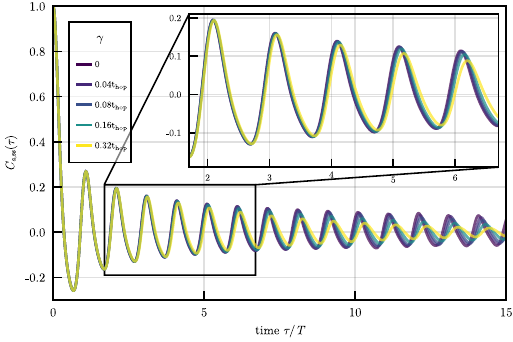}
	\caption{Steady state correlation function $C_{a, ss}(\tau)$ as a function of correlation time $\tau$ in units of period $T$ corresponding to maximum group velocity for different dissipation rates $\gamma$. $\sigma = 0.1N$, $N=200$.}
	\label{fig:correlation_func_D_a_gamma_all_n_100}
\end{figure}
We now consider the two-time correlator of the Lindblad operator $\hat{D}_a$:
\begin{equation}
C_a(t,\tau) = \expval{\hat{D}_a^\dagger(t+\tau)\hat{D}_a(t)}.
\end{equation}
To analyze temporal structures in the steady state, we calculate $C_a(t,\tau)$ for $t\rightarrow \infty$. In this limit, we have
\begin{equation}\label{eq:corr_limit_t}
    \lim_{t\rightarrow\infty} C_a(t,\tau) = \expval{\hat{D}_a^\dagger(\tau)\hat{D}_a}_\mathrm{ss} = \tfrac{1}{N}\mathrm{Tr}\left\{\hat{D}_a^\dagger(\tau)\hat{D}_a\right\},
\end{equation}
where in the last step, we have used the fact that the steady state is a maximally mixed state corresponding to infinite temperature. Taking the limit of $\tau\rightarrow\infty$, in Eq.~\eqref{eq:corr_limit_t}, the operators decorrelate, yielding
\begin{equation}
    \lim_{\tau \rightarrow \infty} \expval{\hat{D}_a^\dagger(\tau)\hat{D}_a}_\mathrm{ss} = \expval{\hat{D}_a}_\mathrm{ss}^2 = \tfrac{1}{N^2} \mathrm{Tr}\left\{\hat{D}_a\right\}^2,
\end{equation}
where we have used that $\hat{D}_a$ is Hermitian.
We construct the normalized correlation function
\begin{equation}
    C_{a, ss}(\tau) = \frac{\expval{\hat{D}_a(\tau)\hat{D}_a}_\mathrm{ss} - \expval{\hat{D}_a}_\mathrm{ss}^2}{\expval{\hat{D}_a(0)\hat{D}_a}_\mathrm{ss} - \expval{\hat{D}_a}_\mathrm{ss}^2}.
\end{equation}
With this definition, $C_{a, ss}(0)=1$ and $C_{a, ss}(\tau \rightarrow\infty)=0$.

The correlation function $C_{a, ss}(\tau)$ is shown in Fig.~\ref{fig:correlation_func_D_a_gamma_all_n_100}(a) for various dissipation rates $\gamma$.
We observe that correlations in $C_{a, ss}(\tau)$ are present even in the steady state with a damped periodicity close to the period $T$ corresponding to maximum group velocity $v_g = 2\mathrm{t}_\mathrm{hop}$. 
The quantum trajectory method provides an intuitive interpretation of this non-trivial behavior: 
Due to subsequent Lindblad operators acting on the system, the wave function becomes localized in space, asymptotically approaching a state centered around $\ket{k}=\ket{\pm \frac{\pi}{2}}$. We observe leading dynamics in each quantum trajectory where the localized wave function moves around the ring with a velocity close to the maximum group velocity. As a change of direction of the wave function in individual quantum trajectories is not forbidden by symmetries and does occur randomly on a characteristic time scale, the steady state correlation function in $C_{a,\mathrm{ss}}(\tau)$ decays eventually.

\subsection{Probability Density Current}
\begin{figure}[tb]
	\centering
	\includegraphics[width=\columnwidth]{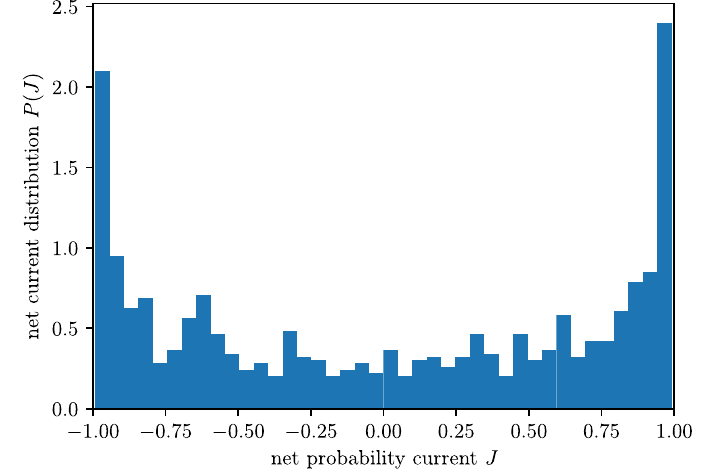}
	\caption{Distribution density of the net probability current $J$ for $N_\text{traj}=1000$ trajectories at $t=500\gamma$ with $\ket{ \psi_\text{init}} = \ket{k=0}$. System parameters: $N=100$, $\mathrm{t}_\mathrm{hop}=25\gamma$, $\sigma=0.1N$.}
	\label{fig:bimodal_distribution}
\end{figure}
We consider the current operator $\hat{\jmath}_a$ between two sites $a$ and $a+1$:
\begin{equation}
	\hat{\jmath}_a = -\mathrm{i}(\hat{c}^\dagger_{a+1} \hat{c}_a - \hat{c}_a^\dagger \hat{c}_{a+1})
\end{equation}
Using the relation $\text{Im}(z) = -\mathrm{i}(z-z^*)/2$, the probability current of state $\ket{\psi}$ at site $a$ is given by
\begin{equation}
\expval{\hat{\jmath}_a} = 2\text{Im}\left(\expval{\hat{c}_a^\dagger \hat{c}_{a+1}}{\psi}\right).
\end{equation}
Summing over all sites we obtain the net probability current $J(t)$ as a function of time:
\begin{equation}
J(t) = \sum\limits_{a=1}^N  \text{Im}\left(\expval{\hat{c}_a^\dagger \hat{c}_{a+1}}\right)
\end{equation}
In the steady state, the probability current expectation value vanishes, characterizing an equilibrium. To see this, we calculate the current operator in quasi-momentum representation in which it takes the form
\begin{equation}
	\expval{\hat{\jmath}_a}{k} = 2\sin(k).
\end{equation}
Calculating the expectation value $\expval{\hat{\jmath}_a}$ in the steady state, we obtain
\begin{equation}
\begin{split}
	\expval{\hat{\jmath}_a}_\text{ss} = \Tr \{\hat{\jmath}_a\hat{\rho}_\text{ss}\} &= \frac{1}{N} \sum\limits_k \expval{\hat{\jmath}_a}{k} \\
 &= \frac{1}{N} \sum\limits_k 2\sin(k) = 0.
 \end{split}
\end{equation}

In the steady state, no spatial direction to which the wave-function moves is preferred, and therefore, we observe a symmetric bimodal distribution in the net probability current in the quantum trajectories at the steady state which we show in Fig. \ref{fig:bimodal_distribution}.

\subsection{Inverse Participation Ratio}
\begin{figure}[b]
    \centering
    \includegraphics[width=\columnwidth]{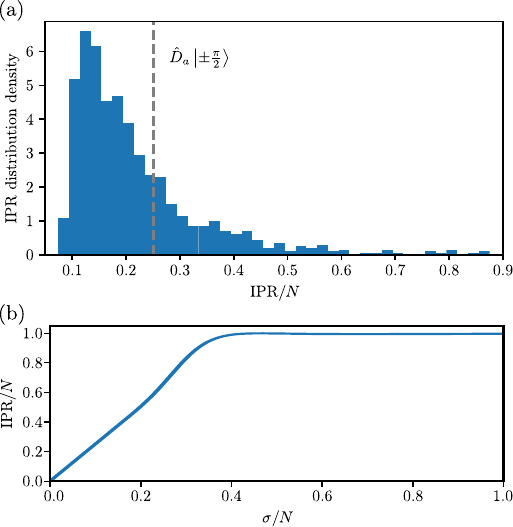}
    \caption{(a) Distribution density of the IPR for $N_\text{traj}=1000$ trajectories at $t=500\gamma$ for an initial state $\ket{\psi_{\text{init}}} = \ket{k=\pi/2}$. System parameters: $N=100$, $\sigma=0.1N$, $\mathrm{t}_\mathrm{hop}=N\gamma/4$. (b) IPR as function of the width $\sigma$ of the amplitude function $h_a(b)$, associated with Lindblad operator $\hat{D}_a$ for a state $\hat{D}_a\ket{k}$.}
    \label{fig:IPR}
\end{figure}
To analyze the degree of \emph{localization} of the wave function in individual quantum trajectories, we define the inverse participation ratio in spatial basis~\cite{Kramer1993}:
\begin{equation}
\begin{split}
&\text{IPR}(\ket{ \psi}) \\ &= \frac{1}{\sum\limits_i |\bra{i}\ket{ \psi}|^{4}} = \begin{cases} 
N & \text{for $\ket{\psi}$ fully extended} \\ 
1 &  \text{for $\ket{\psi}$ fully localized}
\end{cases},
\end{split}
\end{equation} where $\{\ket{i}\}$ is the spatial basis of the system. The system naturally exhibits localization due to the subsequent quantum jumps at different sites as shown in Fig. \ref{fig:IPR}(a).

The IPR for a state with a Lindblad operator $\hat{D}_a$ acting on an arbitrary quasi-momentum state is given by
\begin{equation}
	\mathrm{IPR}(\hat{D}_a\ket{k}) = 1/{\sum\limits_{b} h_a^4(b)},
\end{equation} and thus depends on the width $\sigma$ of the amplitude function $h_a(b)$, as shown in Fig. \ref{fig:IPR}(b).

\subsection{Time-Domain Frequency Analysis}
To analyze the involved frequencies in the quantum trajectory dynamics, we look at the position of the peak amplitude of the localized wave-function. We parameterize the position of the peak amplitude by a discrete angle $\phi_j =  (2\pi j/N \mod N)$ on the ring. Projecting the angle onto the y-axis, $y(t)=\sin(\phi(t))$,  we obtain a quasi-periodic signal.
\begin{figure}[t]
\centering
\includegraphics[width=\columnwidth]{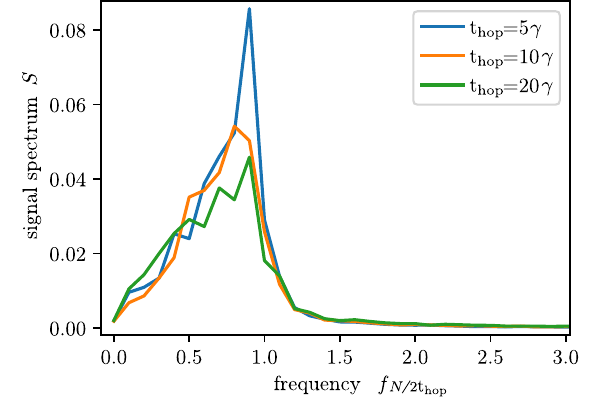}
\caption{Power spectral density $S$ for single quantum trajectories in the long-time limit for different hopping amplitudes $\mathrm{t}_\mathrm{hop}$.}
\label{fig:movingpeakamplitude}
\end{figure}
To quantify involved frequencies in the motion of the peak amplitude position around the ring, we compute the power spectral density, defined by a discrete Fourier transform in the time domain~\cite{Priestley1982}:
\begin{equation}
	S(\omega) = \frac{\Delta t^2}{T} \left|  \sum\limits_{j=1}^{N_t} y_j e^{-\mathrm{i}\omega j\Delta t}\right|^2,
\end{equation}
where $N_t$ is the total number of timesteps for the considered quantum trajectory and $y_j$ is the projection on the y-axis for the $j$-th timestep.

Computing the power spectral density in the long-time limit of individual quantum trajectories, we observe a peak signal at the frequency corresponding to the maximal group velocity of the system Hamiltonian $v_g = \pm 2 \mathrm{t}_\mathrm{hop}$, as shown in Fig.~\ref{fig:movingpeakamplitude}.

\section{Conclusion}
\label{sec:conclusions}
In this work, we have considered a measurement-induced clock in an open quantum system of non-interacting electrons on a periodic lattice, circumventing the requirement of external drives or many-body interactions.

(i) Individual trajectories exhibit wave-function localization due to successive quantum jumps of Lindblad operators modeling continuous quasi-local position measurements of the system. In the steady state, the wave function is found symmetric around quasi-momentum states $\ket{k} = \ket{\pm\frac{\pi}{2}}$, corresponding to the maximum group velocity $v_g = \pm 2\mathrm{t}_\mathrm{hop}$. The interplay between the dissipation rate $\gamma$ and the amplitude function's width $\sigma$ significantly dictates the timescale of unidirectional motion of the wave packet.

(ii) We observe a damped periodicity of two-time expectation values in the steady-state, matching the periodicity of maximum group velocity. The oscillations are sustained for a time scale associated with the dissipation rate $\gamma$ and hopping amplitude $\mathrm{t}_\mathrm{hop}$.
In the steady state, the net probability density current admits a bimodal distribution, indicating the motion of the particle clock-wise or counter-clock-wise around the ring.
Furthermore, using a time-series analysis of the peak-amplitude position of the wave function, we reveal a dominant frequency associated with the particle's motion.

This decaying time-periodic structure, exhibited in the steady state of this chain of non-interacting electrons with measurement-induced interaction, is broadly applicable across a wide parameter range, indicating its robustness and generality.
These results may have potential implications for the design of robust quantum clocks.

\acknowledgements

D.S. acknowledges support by the Swiss National Science Foundation through Projects No. 200020\_185015 and 200020\_215172. S.K. acknowledges partial support through the Deutsche Forschungsgemeinschaft (DFG, German Research Foundation) via Research Unit FOR 2414 under project number 277974659, and by the National Science Foundation under Grants No. NSF PHY-1748958 and PHY-2309135.

\appendix

\section{Derivation of the Liouvillian Equations of Motion}
\label{app:LiouvillianEOM}
Here, we give a detailed derivation of the Liouvillian equation of motion in Eq.~\eqref{eq:Lindblad_final} in the main text.
The master equation for the density matrix $\hat{\rho}$ is given by [c.f. Eq.~\eqref{eq:Linbdblad_a} in the main text]
\begin{equation}
\begin{split}
\frac{\text{d}}{\text{d}t} \hat{\rho}
&=-\mathrm{i}[\hat{H}_\text{TB},\hat{\rho}] + \sum\limits_{a=1}^N\gamma_a\mathcal{D}[\hat{D}_a](\hat{\rho}),
\end{split}
\end{equation}
where $\hat{H}_\text{TB}$ is the tight-binding Hamiltonian of the system, defined in Eq.~\eqref{eq:Hamiltonian} and $\hat{D}_a$ are the Lindblad operators, defined in Eq.~\eqref{eq:D_a_Fourier}.
The dissipator $\mathcal{D}[\hat{A}]$ of a Lindblad operator $\hat{A}$ is defined by
\begin{equation}
    \mathcal{D}[\hat{A}](\hat{\rho}) \equiv \hat{A}\hat{\rho}\hat{A}^\dagger - \frac{1}{2} \hat{A}^\dagger\hat{A}\hat{\rho} - \frac{1}{2}\hat{\rho}\hat{A}^\dagger \hat{A}.
\end{equation}

We express the Lindblad master equation as a super-operator $\mathcal{L}$ acting on the system's density matrix $\hat{\rho}$:
\begin{equation}
\frac{\text{d}}{\text{d}t} \rho_{kk'} \ketbra{k}{k'}= \sum\limits_{\alpha\beta} \mathcal{L}_{ kk'\alpha\beta}\rho_{\alpha\beta}\ket{\alpha}\bra{\beta}.
\end{equation}
To calculate the different matrix elements of the Liouvillian, we first look at the unitary part of Eq.~\eqref{eq:Linbdblad_a}:
\begin{equation}
\begin{split}
&-\mathrm{i}[\hat{H}_\text{TB},\ketbra{k}{k'}] =-\mathrm{i}\left[\sum\limits_\alpha \varepsilon_\alpha \hat{c}_\alpha^\dagger \hat{c}_\alpha,\ketbra{k}{k'}\right] \\
&=-\mathrm{i}\left(\sum\limits_\alpha \varepsilon_\alpha {{}\hat{c}_\alpha}^\dagger \hat{c}_\alpha \ketbra{k}{k'}
- \ketbra{k}{k'} \sum\limits_\alpha \varepsilon_\alpha {{}\hat{c}_\alpha}^\dagger\hat{c}_\alpha\right) \\
&=-\mathrm{i}\left( \varepsilon_k \ketbra{k}{k'} 
- \ketbra{k}{k'} \varepsilon_{k'} \right) \\
&=-\mathrm{i}\left( \varepsilon_k - \varepsilon_{k'} \right) \ketbra{k}{k'}
\end{split}
\end{equation}
It is important to note that the Lindblad operators $\hat{D}_a$ are Hermitian, and hence the dissipator reduces to
\begin{equation}
    \mathcal{D}[\hat{D}_a](\hat{\rho}) \equiv \hat{D}_a\hat{\rho}\hat{D}_a - \frac{1}{2} \hat{D}_a^2\hat{\rho} - \frac{1}{2}\hat{\rho}\hat{D}_a^2.
\end{equation}
Considering the first part of the dissipator, we compute
\begin{equation}
\begin{split}
&\hat{D}_a\ketbra{k}{k'}\hat{D}_a \\ &= 
\frac{1}{N}\left(\sum\limits_{\alpha,\alpha'} \hat{c}_{\alpha}^\dagger \hat{c}_{\alpha'} \eta(\alpha-\alpha')e^{-\mathrm{i}a(\alpha-\alpha')}\right) 
\ketbra{k}{k'}\\
&\times \left(\sum\limits_{\beta,\beta'} \hat{c}_{\beta}^\dagger \hat{c}_{\beta'} \eta(\beta-\beta')e^{-\mathrm{i}a(\beta-\beta')}\right)\\
&=\frac{1}{N}\left(\sum\limits_{\alpha,\alpha'} \hat{c}_{\alpha}^\dagger \delta_{\alpha'k}\eta(\alpha-\alpha')e^{-\mathrm{i}a(\alpha-\alpha')}\right) 
\ketbra{\Omega}{\Omega}\\
&\times \left(\sum\limits_{\beta, \beta'} \delta_{\beta k'} \hat{c}_{\beta'} \eta(\beta-\beta')e^{-\mathrm{i}a(\beta-\beta')}\right)\\
&=\frac{1}{N}\left(\sum\limits_{\alpha} \eta(\alpha-k)e^{-\mathrm{i}a(\alpha-k)}\ket{\alpha}\right)\\ &\times 
\left(\bra{\beta'}\sum\limits_{\beta'} \eta(k'-\beta')e^{-\mathrm{i}a(k'-\beta')}\right)\\
&=\frac{1}{N}\sum\limits_{\alpha, \beta}  \eta(\alpha-k)\eta(k'-\beta) e^{-\mathrm{i}a(\alpha-\beta+k'-k)}\ket{\alpha}\bra{\beta},
\end{split}
\end{equation}
where $\ket{\Omega}$ is the vacuum state.

For $\hat{D}_a^2\hat{\rho}$, we get
\begin{equation}
\begin{split}
&\hat{D}_a^2\ketbra{k}{k'}\\
&=\frac{1}{N}\sum\limits_{\alpha, \alpha',\beta,\beta'} \Big(\hat{c}_{\alpha}^\dagger \hat{c}_{\alpha'}\hat{c}_{\beta}^\dagger \hat{c}_{\beta'} \eta(\alpha-\alpha')e^{-\mathrm{i}a(\alpha-\alpha')}\\
&\qquad\qquad\qquad\times\eta(\beta-\beta')e^{-\mathrm{i}a(\beta-\beta')}\ketbra{k}{k'}\Big)\\
&=\frac{1}{N}\sum\limits_{\alpha, \alpha',\beta,\beta'}\Big( \delta_{\alpha' \beta}\delta_{\beta' k} \eta(\alpha-\alpha')\eta(\beta-\beta')\\
&\qquad\qquad\qquad\times e^{-\mathrm{i}a(\alpha-\alpha'+\beta-\beta')}\ketbra{\alpha}{k'}\Big) \\
&=\frac{1}{N}\sum\limits_{\alpha,\beta} \eta(\alpha-\beta)\eta(\beta-k)e^{-\mathrm{i}a(\alpha-k)}\ket{\alpha}\bra{k'}.
\end{split}
\end{equation}
Analogous for $\ketbra{k}{k'}\hat{D}_a^2$:
\begin{equation}
\begin{split}
&\ketbra{k}{k'}\hat{D}_a^2\\
&=\frac{1}{N}\sum\limits_{\alpha, \alpha',\beta,\beta'} \Big( \ketbra{k}{k'} \hat{c}_{\alpha}^\dagger \hat{c}_{\alpha'}\hat{c}_{\beta}^\dagger \hat{c}_{\beta'}\\
&\qquad\qquad\qquad\times \eta(\alpha-\alpha')e^{-\mathrm{i}a(\alpha-\alpha')}\\
&\qquad\qquad\qquad\times \eta(\beta-\beta')e^{-\mathrm{i}a(\beta-\beta')}\Big)\\
&=\frac{1}{N}\sum\limits_{\alpha, \alpha',\beta,\beta'}\Big( \ketbra{k}{\beta'}\delta_{k' \alpha}\delta_{\alpha' \beta} \eta(\alpha-\alpha')\eta(\beta-\beta')\\
&\qquad\qquad\qquad\times e^{-\mathrm{i}a(\alpha-\alpha'+\beta-\beta')} \Big)\\
&=\frac{1}{N}\sum\limits_{\beta,\beta'} \eta(k'-\beta)\eta(\beta-\beta')e^{-\mathrm{i}a(k'-\beta')}\ket{k}\bra{\beta'}\\
&=\frac{1}{N}\sum\limits_{\alpha,\beta} \eta(k'-\alpha)\eta(\alpha-\beta)e^{-\mathrm{i}a(k'-\beta)}\ket{k}\bra{\beta},
\end{split}
\end{equation}
where in the last step, we have just renamed variables. Thus, the Lindblad master equation is explicitly given by Eq.~\eqref{eq:Lindblad_final}:
\begin{equation}
\begin{split}
&\mathcal{L}[\rho_{k k'}]= \\&-\mathrm{i}\left( \varepsilon_k - \varepsilon_{k'} \right) \rho_{kk'} \\&+ \frac{1}{N}\sum\limits_a \gamma_a  \Bigg\{ \sum\limits_{\alpha,\beta}\Big[\eta(\alpha-k)\eta(k'-\beta)\\
&\qquad \times e^{-\mathrm{i}a(\alpha-\beta+k'-k)}\rho_{\alpha\beta}\\
&\qquad \qquad -\frac{1}{2} \eta(\alpha-\beta)\eta(\beta-k)e^{-\mathrm{i}a(\alpha-k)}\rho_{\alpha k'} \\
&\qquad- \frac{1}{2}\eta(k'-\alpha)\eta(\alpha-\beta)e^{-\mathrm{i}a(k'-\beta)}\rho_{k \beta}\Big]\Bigg\}.
\end{split}
\end{equation}

In the case of all Lindblad operators acting on the system with equal dissipation rates, i.e. $\gamma_a = \gamma = \text{const.}$, the above equation reduces to [c.f Eq.~\eqref{eq:Lindbladequaldissipation} in the main text]:
\begin{equation}
\begin{split}
\mathcal{L}[\rho_{k k'}] &= \left[-\mathrm{i}\left( \varepsilon_k - \varepsilon_{k'} \right)-\gamma\right]\rho_{k k'} \\
&+ \gamma\sum\limits_{\alpha\beta}\eta(\alpha-k)\eta(k'-\beta)\delta_{\alpha-\beta, k-k'}\rho_{\alpha\beta}.
\end{split}
\end{equation}

It is straightforward to check that the above equation of motion for the matrix elements $\rho_{kk'}$ preserves trace and Hermiticity of $\hat{\rho}$.

\section{Derivation of the Exact result}
\label{app:exactresult}

Here, we prove that for an initially mixed density matrix with $\hat{\rho}_0 = \sum_k \rho_{k k}\ketbra{k}{k}$, the time evolved density matrix remains diagonal and real at all times. The time-evolution of the diagonal elements is given by [c.f Eq.~\eqref{eq:exact_result} in the main text]
\begin{equation}\label{eq:exact_result2}
\frac{\text{d}}{\text{d}t} \rho_{kk} = \mathcal{L}[\hat{\rho}] =\gamma  \left(\sum\limits_{\alpha=1}^N\eta^2(\alpha-k) \rho_{\alpha \alpha}-\rho_{k k}\right),
\end{equation}

{\bfseries\sffamily Proof:} The unitary term in the Liouvillian is only nonzero for $k\neq k'$ and thus zero at all times for any initial mixed state. The Liouvillian thus becomes:
\begin{equation}
\begin{split}
\mathcal{L}[\hat{\rho}] &=\frac{1}{N}\gamma\sum\limits_{a=1}^N \left\{ \sum\limits_{\alpha, \beta}\left[\eta(\alpha-k)\eta(k'-\beta) e^{-\mathrm{i}a(\alpha-\beta+k'-k)}\rho_{\alpha \beta}\right.\right.\\
&\qquad-\frac{1}{2} \eta(\alpha-\beta)\eta(\beta-k)e^{-\mathrm{i}a(\alpha-k)}\rho_{\alpha k'} \\
&\left. \left.\qquad- \frac{1}{2}\eta(k'-\alpha)\eta(\alpha-\beta)e^{-\mathrm{i}a(k'-\beta)}\rho_{k,\beta}\right]\right\}
\end{split}
\end{equation} By assumption of the initial state, only diagonal matrix elements contribute to the Liouvillian:
\begin{equation}
\begin{split}
\mathcal{L}[\hat{\rho}] &=\frac{1}{N}\gamma\sum\limits_{a=1}^N  \left\{ \sum\limits_{\alpha, \beta}\left[\eta(\alpha-k)\eta(k'-\alpha) e^{-\mathrm{i}a(k'-k)}\rho_{\alpha \alpha}\right.\right.\\
&\qquad-\frac{1}{2} \eta(k'-\beta)\eta(\beta-k)e^{-\mathrm{i}a(k'-k)}\rho_{k',k'} \\
&\left. \left.\qquad- \frac{1}{2}\eta(k'-\alpha)\eta(\alpha-k)e^{-\mathrm{i}a(k'-k)}\rho_{k k}\right]\right\}.
\end{split}
\end{equation} Since each of the phase factors are equal, summing over $a$ yields only contributions for $k'=k$:
\begin{equation}
\begin{split}
\mathcal{L}[\hat{\rho}] &=\frac{1}{N}\gamma   \sum\limits_{\alpha, \beta}\big[\eta^2(\alpha-k) \rho_{\alpha \alpha}\\
&-\frac{1}{2}\left(\eta^2(\beta-k)+\eta^2(\alpha-k)\right)\rho_{k k}\big].
\end{split}
\end{equation}
Summing over $\alpha$ and $\beta$, and recalling that \mbox{$\sum_k \eta^2(k-k') = 1$} [See Eq.~\eqref{eq:normalization_h_a}], we are left with Eq.~\eqref{eq:exact_result2}, which proofs the identity of Eq. \eqref{eq:exact_result}. We recognize that the fully mixed state $\hat{\rho} = \sum_k \frac{1}{N}\ketbra{k}{k}$ is a fixpoint of the Lindblad master equation since for this state $\mathcal{L}[\hat{\rho}] = 0$.

\section{Sample Trajectories}
\begin{figure*}
    \centering
    \includegraphics[width=\textwidth]{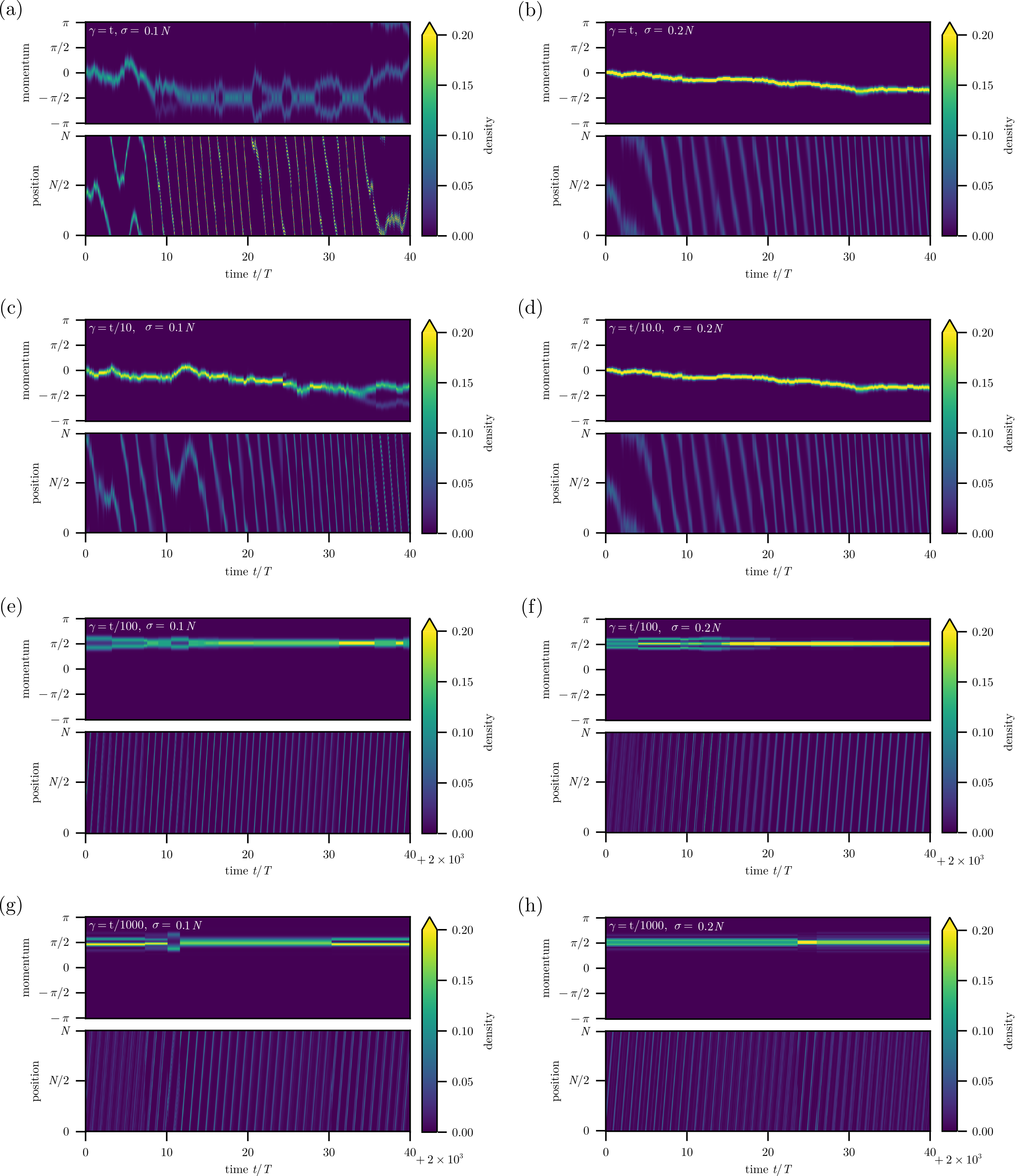}
    \caption{(a-h) Sample trajectories for different system parameters $\gamma, \sigma$ as a function of normalized time $t/T$, where $T$ is the time period associated with maximum group velocity $v_g$. (e-h) For small dissipation rates, $\gamma \ll \mathrm{t}_\mathrm{hop}$, we show the trajectories after the transient dynamics after time $t=2 \times 10^3 T$. All trajectories are initialized with a fully delocalized state, $\ket{\psi_0} \propto \sum_{n=0}^{N-1} \ket{n}$.}
    \label{Fig:Trajectories_Page}
\end{figure*}

In Fig.~\ref{Fig:Trajectories_Page}, we plot various sample trajectories both in quasi-momentum space and position space for different system parameters $\gamma$, and $\sigma$ for a system size of $N=100$.

\vspace*{50em}
\bibliography{bibliography.bib}

\end{document}